\documentclass[conference,a4paper, 11pt]{APSIPA2026}
\IEEEoverridecommandlockouts
\usepackage{amsmath}
\usepackage{graphicx}
\usepackage{multirow}
\usepackage{threeparttable}
\usepackage[backend=biber,style=ieee,]{biblatex}
\usepackage{amssymb,amsmath,bm}
\AtBeginBibliography{\footnotesize}

\usepackage{geometry}
\usepackage{fancyhdr}

\fancypagestyle{firststyle}{
  \fancyhf{}
  \fancyhead[C]{2026 Asia Pacific Signal and Information Processing Association Annual Summit and Conference (APSIPA ASC)}
}

\begin{document}

\title{
BiTSE: 
Binaural Target Speaker Extraction in Noisy Multi-Talker Environments for AR Glass Arrays
}




\author{
\authorblockN{
Selani A. Indrapala and
Wageesha N. Manamperi
}

\authorblockA{
Department of Electronic and Telecommunication Engineering, University of Moratuwa, Sri Lanka \\
E-mail: \textit {\{indrapalasa.20, wageesham\}@uom.lk}}

\thanks{
This work was supported by the Accelerating Higher Education Expansion and Development (AHEAD) operation (Grant No. 6026-LK/8743-LK) 
and
the Telecommunications Regulatory Commission of Sri Lanka (TRCSL) through the Financial Grant for Research and Development 2024.
}
}

\maketitle
\thispagestyle{firststyle}
\pagestyle{empty}

\begin{abstract}
Isolating a desired speech signal in noisy multi-talker conversational scenarios is a key requirement for augmented reality (AR) wearable microphone array systems. In this work, a binaural target speaker extraction (TSE) framework, termed BiTSE, is proposed. It leverages both spatial and temporal cues, specifically the direction-of-arrival (DoA) of the target speaker and corresponding voice activity information, to guide the extraction process. Built upon a binaural signal denoising architecture, our model integrates three key enhancements: (i) a DoA-aware attention mechanism using cyclic positional embeddings, (ii) a timestamp-based masking strategy that utilizes speaker activity to suppress non-target segments, and (iii) a novel two-stage loss optimization strategy that first trains the model for robust denoising and then fine-tunes it to improve perceptual quality. Evaluations on the SPeech Enhancement for Augmented Reality (SPEAR) challenge dataset demonstrate that the proposed BiTSE consistently improves upon conventional approaches, leading to enhanced signal fidelity and perceptual quality.
\end{abstract}


\vspace{-0.2cm}
\section{Introduction}

The cocktail party effect, which refers to the human ability to focus on a single speaker in the presence of multiple overlapping voices, remains a major challenge for 
hearable devices and augmented reality (AR) headsets/glasses \cite{cherry1953some, bregman1994auditory, bronkhorst2000cocktail}. 
Target Speaker Extraction (TSE) focuses on isolating a specific speaker from a mixture using auxiliary information about the target speaker \cite{zmolikova2023neural}. Recent methods leverage diverse cues such as visual information \cite{lin2023av, sato2021multimodal}, acoustic embeddings \cite{zhang2025multi, liu2023x}, and spatial features \cite{wang2025leveraging, wang2024dual, elminshawi2023beamformer}. However, in reverberant environments, single-channel TSE performance degrades due to reverberation-induced smearing of spectro-temporal structure \cite{elminshawi2023beamformer, chen2017cracking}. To address this, multi-microphone approaches are often used to exploit spatial information, with binaural recordings from two spatially separated microphones providing particularly useful spatial cues for improved separation.
This paper presents a neural network-based approach to binaural TSE for AR glasses.

Among the cues used in TSE, spatial information is especially important in multi-microphone settings, where direction-of-arrival (DoA) information is often combined with time–frequency representations for improved performance \cite{wang2024study, gu2021complex}. In addition to cue design, various architectures have been explored, including CNN- and RNN-based models \cite{sinha2021speaker, delcroix2021speaker}, as well as more recent transformer and self-attention approaches \cite{choi2025multichannel, meng2024binaural}. These attention-based methods have shown strong effectiveness in speech enhancement and source separation tasks \cite{tokala2024binaural, kong2022speech}.

TSE involves two key objectives: suppressing interfering speakers and noise, while preserving the perceptual quality of the extracted speech. Prior work has explored loss functions that jointly optimize noise reduction and perceptual fidelity \cite{hernandez2024interaural, kolbaek2020loss}. In particular, the Binaural Complex Convolutional Transformer Network (BCCTN) \cite{tokala2024binaural} improves both signal quality and spatial preservation in binaural speech enhancement, motivating its extension to TSE. However, binaural TSE is more challenging than speech enhancement, as it requires selectively extracting a target speaker from competing sources while maintaining spatial consistency.

\begin{figure*}[t]
\centerline{\includegraphics[width=0.65\textwidth]{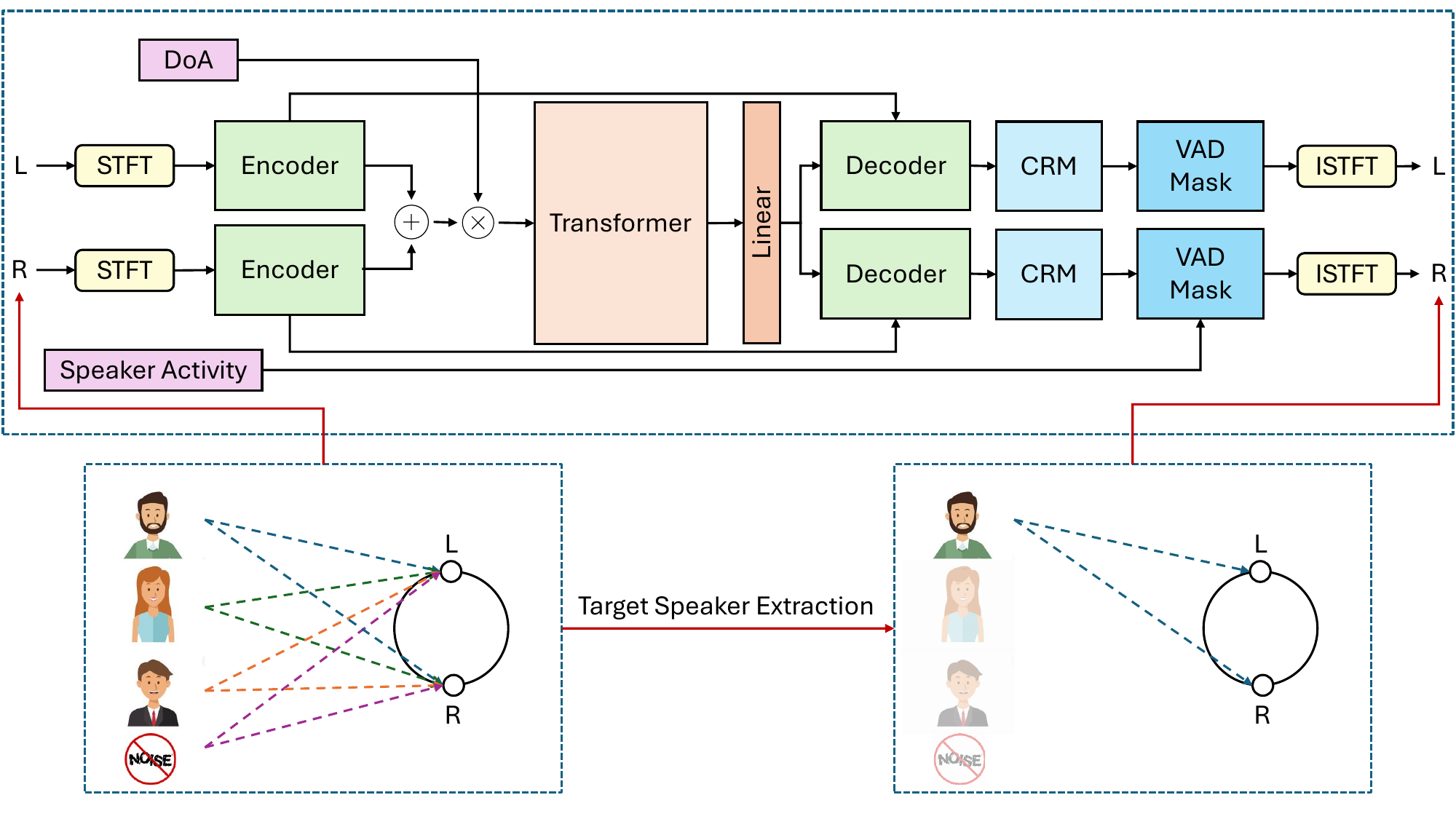}}
\caption{Overview of the proposed BiTSE model. The system operates on binaural microphone inputs containing background noise and competing speakers. For each input channel, the STFT is computed separately and processed through the BCCTN. A DoA embedding is incorporated to guide the separation process. The model estimates a complex ratio mask (CRM), which is further refined using a VAD–based masking strategy. Finally, the inverse STFT is applied to reconstruct the denoised left and right channel signals.}
\label{Fig:Model_Archi}
\end{figure*}

The SPeech Enhancement for Augmented Reality (SPEAR) challenge \cite{tourbabin2023spear} comprises noisy speech recordings captured using a microphone array mounted on AR glasses in a realistic restaurant-like acoustic environment. The dataset is particularly challenging due to strong background noise and multiple competing speakers. Consequently, the target source direction is assumed to be known for the challenge. Apart from this assumption, most participating algorithms rely on pre-measured steering vectors or array responses \cite{tourbabin2023spear}, including the conventional baseline, the Minimum Variance Distortionless Response (MVDR) beamformer \cite{hafezi2023subspace}.

In this paper, BiTSE, a deep learning-based binaural TSE algorithm, is proposed, where BCCTN \cite{tokala2024binaural} is utilized to extract the desired speech from multiple interference speakers in noisy, reverberant environments using both the target source direction and ground-truth voice activity detection (VAD) as prior information. This work explores the integration of spatial conditioning, target activity guidance, and perceptually-oriented optimization within a binaural complex-domain transformer framework for robust TSE under adverse acoustic conditions. The main aspects investigated in this work are: 
(i) DoA-informed conditioning to incorporate spatial cues within binaural target speaker extraction; and 
(ii) speaker activity-based masking to emphasize target-active segments during extraction.
We also propose a novel two-stage loss optimization strategy that explicitly addresses denoising and preservation of perceptual quality, enhancing both the accuracy and naturalness of extracted speech. 

The rest of the paper is organized as follows. Sections~\ref{sec:azimuth_attention} through \ref{sec:loss_function} present the proposed azimuth-aware attention mechanism, speaker activity-based masking strategy, and two-stage loss optimization, respectively. In Section~\ref{sec:Experiments}, we evaluate the proposed BiTSE algorithm on three distinct labeled datasets: D2, D3, and D4, from the SPEAR challenge \cite{tourbabin2023spear}, and compare its performance against BCCTN variants and the classical MVDR beamformer \cite{hafezi2023subspace}, where BiTSE consistently surpasses state-of-the-art methods.

\vspace{-0.1cm}
\section{Problem Formulation}
\vspace{-0.1cm}

The objective is to extract a target speaker from a binaural mixture under low Signal-to-Noise Ratio (SNR), overlapping speech, and intermittent silence. 

For each channel $c \in \{L, R\}$, where L and R denote the left and right channel respectively, the time-domain mixture is modeled as:  
\vspace{-0.2cm}
\begin{equation}
x_c(t) = \sum_{k=1}^{K} \hat{s}_{c,k}(t) + n_c(t), 
\end{equation}
where $\hat{s}_{c,k}(t)$ is the contribution of the $c^{th}$ channel of speaker $k$, and $n_c(t)$ is additive noise at the $c^{th}$ channel. In the short-time Fourier transform (STFT) domain $x_c(t)$ can be defined as,  
\vspace{-0.2cm}
\begin{equation}
X_c(t, f) = \sum_{k=1}^K H_{c,k}(f, \theta_k) S_k(t, f) + N_c(t, f),
\end{equation}
where $H_{c,k}(f, \theta_k)$ denotes the acoustic transfer function between the $k^{th}$ source and $c^{th}$ microphone channel for a direction-of-arrival (DoA) $\theta_k$.  

Given the binaural mixture $X_c(t,f)$ for $c \in \{L,R\}$, and an auxiliary clue $C_g$, which is used here as the target DoA, the target speaker extraction (TSE) model estimates the target signal at the $c^{th}$ channel as:  
\vspace{-0.2cm}
\begin{equation}
\hat{S}_{c,\text{target}} = \mathrm{TSE}(X_c(t,f), C_g; \boldsymbol{\Theta}),
\vspace{-0.2cm}
\end{equation}
where $\boldsymbol{\Theta}$ denote the model parameters. 

The aim of this paper is to recover the target source signal $\hat{S}_{c,\text{target}}(t,f)$ arriving from azimuth direction of $\theta_{\text{target}}$, using both the microphone recordings, $X_c(t, f)$, and a voice activity detection (VAD) label, $v(t) \in \{0,1\}$, which indicates target activity of the desired source. 

\vspace{-0.2cm}
\section{Model architecture}
\vspace{-0.1cm}

In this section, we present the proposed binaural TSE (BiTSE) architecture in detail. The model is designed to leverage binaural cues, spatial embeddings, and temporal information to achieve robust target speaker extraction. The proposed model, as depicted by Fig.~\ref{Fig:Model_Archi}, builds upon the Binaural Complex Convolutional Transformer Network (BCCTN) architecture\footnote{While BCCTN is designed for single speaker binaural speech enhancement, BiTSE extends it to target speaker extraction by incorporating DoA-informed attention, speaker activity–based masking, and a two-stage loss optimization strategy.} \cite{tokala2024binaural}, which processes complex-valued binaural spectrograms through a symmetric dual-branch design for the left and right channels. The model retains the complex convolutional encoder-decoder structure of BCCTN to extract spectral-spatial representations from each channel. 


\begin{figure}[h!]
\centerline{\includegraphics[width=1.0\columnwidth, trim=0 120 0 0, clip]{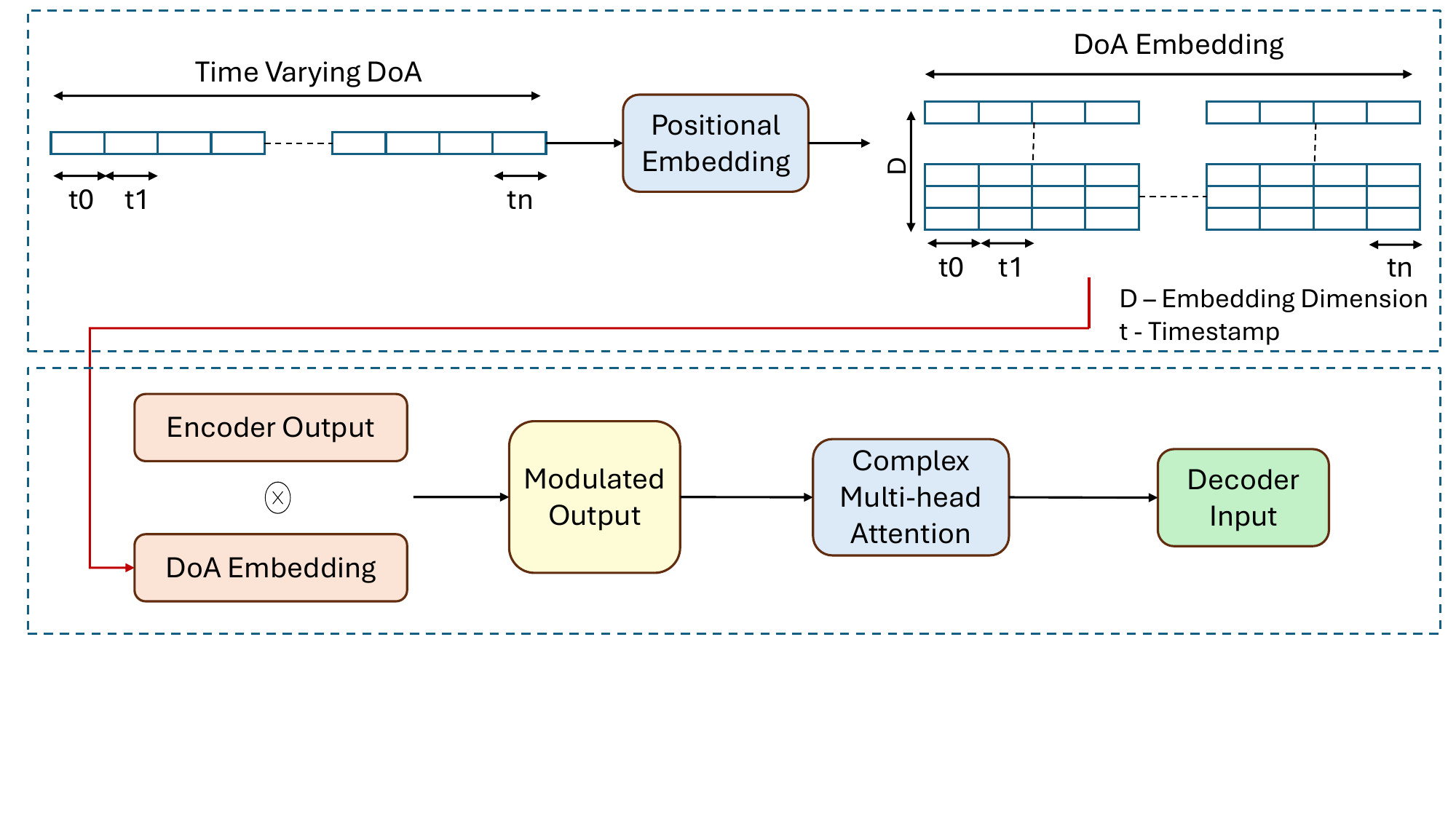}}
\caption{Overview of the azimuth-aware self-attention module in the transformer block of Fig. \ref{Fig:Model_Archi}. The time-varying DoA trajectory is first converted into a DoA embedding, which modulates the encoder output. The modulated features are then processed by a complex-valued multi-head self-attention module to produce the input for the decoder.}
\label{Fig:Att_archi}
\end{figure}

\vspace{-0.3cm}
\subsection{Azimuth-Aware Self Attention} \label{sec:azimuth_attention}
\vspace{-0.1cm}

In this work, we use the azimuth of the target speaker as a clue for TSE. The DoA information is first temporally aligned with the STFT time resolution by averaging over each STFT frame's corresponding time window. Secondly, to incorporate azimuthal information as a positional prior, we adopt a \textit{cyclic positional encoding} strategy as in \cite{choi2025multichannel}. Specifically, for an azimuth angle $\phi$, the embedding vector $\mathrm{PE}_{\mathrm{cyc}}(\phi) \in \mathbb{R}^{D}$ is defined as,
\begin{equation}
\mathrm{PE}_{\mathrm{cyc}}(\phi,2j)=
\sin\left(
\sin(\phi)\cdot\frac{\alpha}{10000^{2j/D}}
\right)
\end{equation}
\begin{equation}
\mathrm{PE}_{\mathrm{cyc}}(\phi,2j+1)=
\sin\left(
\cos(\phi)\cdot\frac{\alpha}{10000^{2j/D}}
\right)
\end{equation}
where $j \in [0, D/2)$, $D$ denotes the embedding dimension, and $\alpha$ is a scaling factor controlling the angular variation of the encoding. In our implementation, $\phi$ is represented in radians, and we empirically set $D=40$ and $\alpha=20$, following \cite{choi2025multichannel}. The resulting embedding is L2-normalized and broadcast along the temporal dimension to match the encoder feature resolution. The resulting azimuth embeddings are then passed through a fully connected layer followed by a PReLU activation. Next, the azimuth embeddings are element-wise multiplied with the concatenated encoder output (where the concatenation operation is represented by the $"+"$ symbol in Fig. \ref{Fig:Model_Archi}), allowing spatial modulation. Finally, the azimuth-aware features are processed by a complex-valued self-attention transformer module. The architecture for the azimuth aware attention is depicted in Fig.~\ref{Fig:Att_archi}. The output of the azimuth-aware attention module is then passed to the mask estimation stage to selectively extract the timestamps where the target speaker is active.

\vspace{-0.1cm}
\subsection{Mask Estimation} \label{sec:timestamp_masking}
\vspace{-0.1cm}

The proposed model employs a two-stage masking strategy to isolate the target speaker while suppressing interference and noise. First, a complex ratio mask (CRM) $\mathbb{M}_c(t,f)$ is estimated for each channel $c \in \{L,R\}$ and applied to the mixture STFT as $\hat{S}_c(t, f) = \mathbb{M}_c(t, f) \cdot X_c(t, f)$, where $X_c(t,f)$ is the observed binaural mixture. To further enforce temporal consistency, a VAD-derived binary mask $\mathbb{V}(t) \in \{0,1\}$ is applied, yielding the estimated target source signal at $c^{th}$ channel as $\tilde{S}_c(t, f) = \mathbb{V}(t) \cdot \hat{S}_c(t, f)$. Here we set $\mathbb{V}(t) = 1$, if the target speaker is active in frame $t$, and $\mathbb{V}(t) = 0$ otherwise. Integrating the VAD mask within the model improves TSE performance, particularly in inactive speech regions, where the CRM alone may produce unreliable or irrelevant estimates.

\vspace{-0.2cm}
\subsection{Loss Function} \label{sec:loss_function}
\vspace{-0.2cm}
TSE requires addressing two key objectives: (i) denoising, by suppressing interfering speakers and noise, and (ii) preservation of perceptual quality. To this end, we adopt a two-stage training strategy to jointly optimize both objectives.  

In the first stage, the model is optimized for denoising using energy-based fidelity losses. We consider both SNR and the segmental SNR (SegSNR)\footnote{SegSNR is computed using the publicly available pysepm library.}, defined as:  
\begin{equation}
\mathcal{L}_{\text{SNR}} = -10 \log_{10} \left( \frac{\sum_t s^2(t)}{\sum_t (s(t) - \hat{s}(t))^2 + \epsilon} \right),
\end{equation}
\begin{equation}
\mathcal{L}_{\text{SegSNR}} = -\frac{1}{K} \sum_{k=1}^{K} 10 \log_{10} \left( \frac{\sum_{t \in \mathcal{S}_k} s^2(t)}{\sum_{t \in \mathcal{S}_k} (s(t) - \hat{s}(t))^2 + \epsilon} \right),
\end{equation}
where $s(t)$ and $\hat{s}(t)$ are clean and estimated signals respectively, $\mathcal{S}_k$ is the $k^{th}$ segment, and $\epsilon$ is defined to ensure the stability. SegSNR is computed over segments defined by the STFT analysis window, using a frame length of 1024 samples with a frame shift of 512 samples.

The second stage fine-tunes the model for perceptual and spatial quality using additional loss terms: short-time objective intelligibility (STOI) \cite{taal2011algorithm},
\begin{equation}
\mathcal{L}_{\text{STOI}} = -\frac{\text{STOI}_L + \text{STOI}_R}{2},
\end{equation}
and an interaural phase difference (IPD) loss,  
\begin{equation}
\mathcal{L}_{\text{IPD}} = \frac{1}{N} \sum_{k, \ell} M(k,\ell) \left| \text{IPD}_S(k,\ell) - \hat{\text{IPD}}_S(k,\ell) \right|,
\end{equation}
where $M(k,\ell)$ is a speech mask and $N$ is the number of active time-frequency (TF) bins. Following Tokala \textit{et al.}~\cite{tokala2024binaural}, the IPD is computed as 
\vspace{-0.2cm}
\begin{equation}
\text{IPD}_S(k, \ell) = \arctan\left( \frac{S_L(k, \ell)}{S_R(k, \ell)} \right),
\end{equation}
where $\arctan(\cdot)$ denotes the inverse tangent function used to compute the phase difference between the left and right channels.

The final training objectives are:  
\vspace{-0.3cm}
\begin{equation}
\mathcal{L}_{\text{stage-1}} = \gamma \cdot \mathcal{L}_{\text{SegSNR}},
\end{equation}
\begin{equation}
\mathcal{L}_{\text{stage-2}} = \gamma \cdot \mathcal{L}_{\text{SegSNR}} + \alpha \cdot \mathcal{L}_{\text{IPD}} + \beta \cdot \mathcal{L}_{\text{STOI}},
\end{equation}
where $\gamma$, $\alpha$, and $\beta$ are the weight factors for $\mathcal{L}_{\text{SegSNR}}$, $\mathcal{L}_{\text{IPD}}$, and $\mathcal{L}_{\text{STOI}}$, respectively. Here we utilize a two-stage loss optimization where stage~1 uses a higher learning rate ($\eta_1$) for robust denoising and stage~2 adopts a smaller rate ($\eta_2$) for perceptual refinement. The two-stage optimization strategy first establishes denoising and subsequently refines perceptual quality and spatial consistency without disrupting the learned speech reconstruction capability.

\vspace{-0.2cm}
\section{Experiments} \label{sec:Experiments}
\vspace{-0.2cm}

\begin{table*}[h!]
\centering
\caption{Performance comparison with the baseline method on datasets D2, D3, and D4 using SegSNR and PESQ metrics.}
\label{tab:results}
\resizebox{0.65\textwidth}{!}{%
\begin{tabular}{c|cc|cc|cc}
\hline
\textbf{Model} & \multicolumn{2}{c|}{\textbf{D2}} & \multicolumn{2}{c|}{\textbf{D3}} & \multicolumn{2}{c}{\textbf{D4}} \\
\cline{2-7}
 & SegSNR $\uparrow$ & PESQ $\uparrow$ & SegSNR $\uparrow$ & PESQ $\uparrow$ & SegSNR $\uparrow$ & PESQ $\uparrow$\\
\hline
Unprocessed & -8.69 & 1.45  & -8.67 & 1.46  & -7.34 & 1.21\\
MVDR (6 channels) & -8.27 & 1.40 & -8.32 & 1.46 & -6.65 & 1.20\\
BCCTN & -8.78 & 1.64 & -9.15 & 1.71 & -8.26 & 1.31 \\
BCCTN + DoA & -9.10 & 1.67 & -9.17 & 1.75 & -8.34 & 1.33 \\
Proposed Method & \textbf{-2.57} & \textbf{1.72} & \textbf{-2.57} & \textbf{1.80} & \textbf{-4.05} & \textbf{1.34} \\
\hline
\end{tabular}
}
\end{table*}

\vspace{-0.1cm}
\subsection{Dataset}
\vspace{-0.1cm}

We evaluate our proposed model using the SPeech Enhancement for Augmented Reality (SPEAR) Challenge dataset \cite{tourbabin2023spear}. The dataset consists of a 6-channel audio mixture captured from an augmented reality (AR) glass array. In this work, we focus on channels 5 and 6, corresponding to the binaural left and right ear signals, respectively. In addition to the audio recordings, each file provides DoA trajectories for all active speakers, as well as VAD labels for each speaker, both sampled at 20 Hz.

We leverage ground-truth DoA and VAD annotations from the SPEAR dataset, allowing us to focus on the design and evaluation of the proposed BiTSE architecture by bypassing separate source direction and speaker activity estimation, and isolating the assessment of the extraction and enhancement components.

The dataset is organized into three subsets with increasing acoustic complexity. Dataset 02 (D2) contains audio samples with simulated reverberation and environmental noise. Dataset 03 (D3) extends D2 by incorporating speaker and listener motion to simulate dynamic acoustic scenes. Dataset 04 (D4) represents the most challenging conditions, combining high noise levels, reverberation, head movements, and overlapping speech. All datasets are characterized by low SNR and high competing speaker conditions, making target speaker extraction particularly challenging. In D2 and D3, there are over 38\% of multi speaker segments with the audio containing at least 2 concurrent speakers. In contrast, D4 is dominated by overlapping speech, with 26\%, 43\%, and 29\% of the audio containing 2, 3, and 4 concurrent speakers.

\vspace{-0.2cm}
\subsection{Training Details}
\vspace{-0.1cm}

The proposed model is trained using the two-stage optimization process described in Section~\ref{sec:loss_function}.
We train the initial denoising stage with a learning rate of $\eta_1=0.001$ and employ a lower learning rate of $\eta_2=0.0001$ during the subsequent fine-tuning stage. Both stages are trained for 100 epochs each using the Adam optimizer with a batch size of 8. The SPEAR dataset provides predefined \textit{Main} and \textit{Dev} splits. We define the loss as a weighted combination of denoising, spatial, and intelligibility objectives, with empirically chosen weights \(\gamma = 1\), \(\alpha = 5\), and \(\beta = 1\) for \(\mathcal{L}_{\text{SegSNR}}\), \(\mathcal{L}_{\text{IPD}}\), and \(\mathcal{L}_{\text{STOI}}\), respectively. The mixture signals are downsampled to 16~kHz before being fed into the model. We also filter out the audio snippets to ensure that at least 60\% of the audio is of an active speech segment. This filtering criterion was applied independently within each data split. The short-time Fourier transform (STFT) uses a window size of 1024 samples with a hop size of 512 samples. The model architecture, shown in Fig.~\ref{Fig:Model_Archi}, incorporates 16 attention heads in the azimuth-aware self-attention module with a total of approximately 2.6 million parameters.

\vspace{-0.2cm}
\subsection{Results and Discussion}
\vspace{-0.1cm}

The proposed BiTSE model is evaluated using two quantitative metrics: SegSNR and Perceptual Evaluation of Speech Quality (PESQ) \cite{recommendation2001perceptual}, which respectively measure denoising performance and perceptual quality.

Table~\ref{tab:results} compares the proposed BiTSE with the unprocessed binaural mixture, a 6-channel Minimum Variance Distortionless Response (MVDR) beamformer \cite{hafezi2023subspace} output, the baseline BCCTN denoiser \cite{tokala2024binaural}, and BCCTN augmented with DoA embeddings across datasets D2–D4. The MVDR beamformer uses the 6-channel head-worn microphone array configuration described in \cite{hafezi2023subspace}, consisting of four microphones mounted on eyeglasses and two in-ear microphones. The beamformer weights are computed using the standard MVDR formulation, where the steering vector is obtained from the target DoA and corresponding acoustic transfer functions (ATFs), while the noise covariance matrix is estimated using an exponentially averaged sample covariance matrix.
The unprocessed signals exhibit highly negative SegSNR values and low PESQ scores, reflecting challenging acoustic conditions. The 6-channel MVDR beamformer provides only marginal improvement, particularly in SegSNR, despite utilizing more channels than both the BCCTN and the proposed model. As discussed in \cite{hafezi2023subspace}, this could be due to adaptive beamformers being sensitive to imperfections in target ATFs and DoA estimation, which may introduce target distortion in dynamic multi-speaker environments. The baseline BCCTN shows better perceptual quality performance than beamforming, demonstrating the advantage of learning based binaural denoising. Adding DoA embeddings to BCCTN yields small gains, highlighting the benefit of incorporating spatial cues. However, in both the above model architectures, the improvement is only in the PESQ. 
Alternatively, the proposed BiTSE model outperforms all baseline methods, achieving the greatest improvements across all datasets.
SegSNR improves more than threefold over the unprocessed input, while PESQ consistently increases. While SegSNR values remain negative, the consistent PESQ gains indicate effective suppression of perceptually disturbing noise, with spectral modifications penalized by SegSNR. 

\begin{table}[h!]
\centering
\caption{
Loss function variants
with dataset D2}
\label{tab:loss_ablation}
\resizebox{0.75\columnwidth}{!}{%
\begin{tabular}{c|c|c|c|c}
\hline
\textbf{$\mathcal{L}_{\text{Denoise}}$} & \textbf{$\mathcal{L}_{\text{IPD}}$} & \textbf{$\mathcal{L}_{\text{STOI}}$} & SegSNR $\uparrow$ & PESQ $\uparrow$\\
\hline
 & - & - & -2.50 & 1.66 \\ 
SNR & $\alpha$ = 5 & - & -2.58 & 1.67 \\ 
 & $\alpha$ = 0.1 & - & -2.61 & 1.67 \\ 
 & - & $\beta$ = 5 & -2.86 & 1.67 \\ 
 & - & $\beta$ = 1 & -2.70 & 1.68 \\ 
 & $\alpha$ = 5 & $\beta$ = 1 & -2.72 & 1.67 \\ 
\hline
 & - & - & -2.66 & 1.64 \\ 
SegSNR & $\alpha$ = 5 & - & -2.45 & 1.70 \\ 
 & $\alpha$ = 0.1 & - & -2.47 & 1.70 \\ 
 & - & $\beta$ = 5 & \textbf{-2.32} & 1.68 \\ 
 & - & $\beta$ = 1 & -2.50 & 1.69 \\ 
 & $\alpha$ = 5 & $\beta$ = 1 & -2.57 & \textbf{1.72} \\ 
\hline
\end{tabular}
}
\end{table}

Table~\ref{tab:loss_ablation} presents an ablation study evaluating the model under different loss configurations. When trained with \( \mathcal{L}_{\text{SNR}} \) alone, the model achieves SegSNR of -2.50 dB and PESQ of 1.66, indicating strong global noise suppression. However, adding STOI or IPD losses produces only marginal changes in SegSNR, suggesting that the SNR-based loss encourages overly aggressive global energy minimization, which can inadvertently suppress fine-grained temporal and spectral variations in speech, leading to an oversmoothing effect. In contrast, using \( \mathcal{L}_{\text{SegSNR}} \) alone achieves slightly lower SegSNR (-2.66 dB) and PESQ (1.64). However, combining SegSNR with intelligibility-focused losses (\(\mathcal{L}_{\text{STOI}} \) and \( \mathcal{L}_{\text{IPD}} \)) further improves PESQ to 1.72 and SegSNR to -2.57 dB, demonstrating a better balance between noise suppression and perceptual quality. This indicates that while \( \mathcal{L}_{\text{SNR}} \) yields stronger uniform noise suppression, it tends to oversmooth speech, whereas \( \mathcal{L}_{\text{SegSNR}} \) better preserves temporal and spectral structures. We note that the speaker extraction results can be further enhanced by the use of post-filters to cancel residual noise \cite{Manamperi2022GMMBM,cheong2024postfilter}.

\vspace{-0.1cm}
\section{Conclusions}
\vspace{-0.1cm}


In this work, we propose BiTSE, a binaural target speaker extraction algorithm for multiple talkers in noisy reverberant environments.
The method integrates DoA-informed embeddings, speaker activity-based masking, and a two-stage loss optimization strategy.
Experimental results show that the proposed BiTSE framework improves performance over beamforming and neural network baselines on the SPEAR dataset. 
Future work will involve subjective testing evaluations, extensive comparison with other deep learning models, and investigation of model optimization techniques, such as quantization and model pruning, for real-time deployment on AR glasses.

\vspace{-0.1cm}
\section*{Acknowledgment}
\vspace{-0.1cm}
The authors would like to thank Prof. Thushara~D.~Abhayapala and Prof. Dileeka~Dias for support on this work. Generative AI has been assisting in writing scripts. Grammatical models have been used for grammar aid.

\printbibliography

\end{document}